# Switching mechanism in polar columnar mesophases made of bent-core molecules

Ewa Gorecka[1], Nataša Vaupotič[2,3], Damian Pociecha[1], Mojca Čepič[2] and Jozef Mieczkowski[1]

[1] *Chemistry Department, Warsaw University, Al. Zwirki i Wigury 101, 02-089 Warsaw, Poland*

[2] *Jozef Stefan Institute, Jamova 39, 1000 Ljubljana, Slovenia*

[3] *Faculty of Education, Koroška 160, 2000 Maribor, Slovenia*

The behavior of polar columnar phases made of bent shaped molecules ($B_{1Rev}$ and $B_{1Rev\ Tilted}$) was studied under applied electric field. There are two competing mechanisms of switching in polar $B_{1Rev\ Tilted}$ columnar phase: the collective rotation around the long molecular axis and collective rotation around the tilt cone. Respectively monostable or bi-stable optical switching is observed for antiferroelectric $B_{1Rev\ Tilted}$ phase. The main factor discriminating the type of the switching is a width of the column cross-section.

## Introduction

There are two broad classes of columnar mesophases; phases made of disc-like molecules [1] and phases made of rod-like molecules, in which columns are formed by smectic layer fragments [2]. The columnar phase made of discs becomes polar if the molecules forming columns are chiral and tilted from the axis of the column [3]. In such a phase every column has spontaneous electric polarization vector perpendicular to the column axis. It has been recently shown that also columnar mesophases made of broken layers can have spontaneously ordered dipoles if made of so called *banana* molecules [4-7]. Due to their *bent-shaped* core these molecules cannot rotate freely around the long axis (a line joining the *banana* ends) and the steric hindrances result in ordering of dipole moments. The spontaneous polarization in the broken-layer columnar *banana* phase can be either perpendicular to the column axis (the $B_1$ phase [4,5]) or along column in the $B_{1Rev}$ - type phases (Fig.1a). These two types of structure can be easily recognized from their X-ray patterns as the $B_1$ phase has the non-centered unit cell and $B_{1Rev}$ phase has the centered unit cell [6-8]. The $B_{1Rev}$ – type phase can be built either of non-tilted ($B_{1Rev}$) or tilted ($B_{1Rev\ Tilted}$) smectic layers. It seems that the $B_1$ phase, although pyroelectric, has spontaneous electric polarization that is not re-orientable in an electric field. The sensitivity of the $B_1$ phase to the electric field, observed in some materials, is the effect related to electric field induced phase transition to the $B_2$ phase [9]. This transition is accompanied by a complete re-building of the texture, both the shape of the domains and the birefringence are changed. In the induced $B_2$ phase the switching with the reorientation of the optical director tensor eigen directions is observed.

In this paper we limit the discussion to the $B_{1Rev}$ phases in which, contrary to the $B_1$ phase, spontaneous polarization can be easily re-oriented by the external electric field without destroying the columnar phase structure. In the orthogonal phase molecules can only rotate around their long axes, while in the tilted phase the switching mechanism becomes complex, as it involves two competitive processes: molecular rotation on the tilt cone around the layer normal and rotation around the long molecular axis.

**Experimental**

Most of electrooptic and dielectric studies were performed in commercially available 5 μm thick glass cells (supplied by Linkam) with ITO electrodes and aligning polyimide surfactant layers. For temperature control the heating stage Mettler FP82HT was used. Dielectric permittivity dispersion was measured with Solartron SI1260 Impedance Analyzer. Texture changes under applied electric field (SRS DS345 Function Generator and FLCE A400 Voltage Amplifier) were observed with polarizing microscope Nikon Optiphot2-pol.

**Results**

For the experiment several materials that have orthogonal $B_{1Rev}$ and/or tilted $B_{1Rev\ Tilted}$ phase were chosen (Table 1). The rectangular and oblique crystallographic unit cell for $B_{1Rev}$ and $B_{1Rev\ Tilted}$ mesophase, respectively was confirmed by X-ray studies (Table 1) [6,8]. The studies show that in both phases there are two columns in crystallographic unit cell, and each cross section of the column contains from 10 to 30 molecules, depending on the material. The tilt angle in the $B_{1Rev\ Tilted}$ phase is small, usually between $5^o$ and $15^o$.

In both, orthogonal and tilted phases the current peak was registered showing polar, switchable nature of the phase. The current peak appears above certain threshold field (usually about $15V_{pp}/\mu m$). The spontaneous electric polarization ($P_s$) as high as 500 $nC/cm^2$ was measured. For compounds showing the $B_{1Rev}$ - $B_{1Rev\ Tilted}$ phase transition no anomaly in the $P_s$ value at the phase transition from the orthogonal to the tilted phase was recorded. In the dielectric studies a single, weak ($\Delta\varepsilon \sim 2$) mode was observed in $B_{1Rev}$ – type phases. The relaxation frequency ($f_r$) of the mode follows the Arhenius law, $f_r \sim e^{-\frac{E_a}{kT}}$, with $E_a \sim 145$ kJ $mol^{-1}$. It should be noticed that the dielectric mode in columnar phases is distinctly weaker than in the lamellar antiferroelectric $B_2$ phase [10] having similar value of the spontaneous electric polarization (Fig. 2). The threshold type switching together with low value of static dielectric constant point to the antiferroelectric nature of the phase, although in most cases only a single current peak is registered.

In the orthogonal $B_{1Rev}$ phase switching in the electric field occurs without optical changes. Both switched-on and switched-off states have the same extinction directions when viewed between crossed polarizers, showing that the principle axes of the optical dielectric tensor do not change after applying the electric field. Only a small increase of the birefringence in the electric field was observed: $\Delta n$ raises from 0.14 to 0.15. The observed changes correspond to the rotation of the molecules around their long axes. A similar switching mechanism has been reported for the lamellar, polar, non-tilted (SmAP) phase made of bent-core materials [11].

In the $B_{1Rev\ Tilted}$ phase upon the transition from the $SmAP_R$, uniaxial lamellar phase reported recently in [12] or the columnar orthogonal $B_{1Rev}$ phase small domains are formed, that grow to a hundred of micron size in the course of applying electric field for several minutes. In the ground state, well aligned samples shows two types of domains (Fig. 3b), each type of domain can be brought into light extinction condition by rotating the sample between crossed polarizers. The domains are synclinic and differ by $2\theta$ in the tilt direction. Close examination of the texture show that the boundaries between two opposite-tilt domains are made of optically active regions (Fig. 4), with the optical rotatory power (ORP) across the cell thickness about twice the value of the tilt angle. These are regions in which the opposite-tilt synclinic domains, anchored at the glass surfaces, overlap. It can not be excluded that in the overlapping region material breaks into regular small blocks, twisted from each other [13].

When a low frequency triangular electric field is applied, the synclinic domains do not change their brightness nor shape (Fig. 3a-c), although reversing of polarization is proved by the appearance of the current peak. The same, mono-stable optical switching is observed if dc field is applied in the sequence $+E$, 0, $-E$. Lack of optical changes upon applying the electric field shows that re-orientation of spontaneous polarization direction to the external electric field leaves the optical dielectric tensor eigen directions unchanged. Thus the switching between the antiferroelectric ground state and the ferroelectric field-on state occurs without changing the synclinic phase structure, by molecular rotation around the long axis in half of the columns. However, if the ac square voltage is applied to the same structure or the dc electric field sign is reversed suddenly (in sequence $+E$, $-E$) clear optical changes are observed as the synclinic domains interchange their brightness (Fig. 3c-d). In this case the dielectric tensor re-orients as the adjusting of the polarization direction to the direction of electric field takes place by collective molecular rotation on the tilt cone in all columns. Apparently, in this case switching between two ferroelectric states occurs without restoring the antiferroelectric ground state, as it is quite often observed in antiferroelectric phase made of chiral rod-like molecules.

Rarely, another type of bi-stable optical switching could also be observed in some materials forming the $B_{1Rev\ Tilted}$ phase. When examining texture with circular domains, the disclination defects around which the splay of crystallographic directions takes place, it can be seen that in the ground state extinction brushes are inclined from polarizer directions while under electric field they take position parallel to the direction of polarizers (Fig. 5). This corresponds to the field induced change from the synclinic antiferroelectric structure to the ferroelectric state with the anticlinic orientation of directors in crystallographic unit cell. The anticlinic arrangement of columns is achieved by molecular rotation on the tilt cone in half of the columns.

It should be stressed that switching by rotation around the cone and around the long axis has different impact on the structural chirality [9]. The rotation around the long axis reverses the structural chirality in the layer fragment forming column, while the rotation around the cone preserves the structural chirality [14].

Further complications in the switching mechanism in the $B_{1Rev\ Tilted}$ phase come from surface anchoring conditions of *banana* molecules. Two types of optical response in the $B_{1Rev\ Tilted}$ phase described above were observed in the *book-shelf* geometry, in which polarization vector **p** is perpendicular to the glass plate. However, for the planar alignment of bent shaped molecules (director parallel to the substrate) another energy minimum exists, for the molecules arranged with the *banana* polarization vector **p** parallel to the glass plane, as already discussed for the $B_2$ phase in [15]. The anchoring condition depends strongly on the surface preparation and might change with temperature or the ionic purity of the compound. It has been observed that for $B_{1RevTilted}$ phase it depends also on the electric field characteristic. Applying a square voltage for several minutes induces a texture with the *banana* polarization perpendicular to the glass plane while under triangular voltage the texture in which *banana* polarization is in the glass plane is obtained. The later case implies that the layer normal is tilted in respect to the glass substrates. If *banana* molecules are oriented with polarization parallel to the glass plate the texture with the extinction directions along and perpendicular to the layer normal is observed. For this texture usually no extinction direction changes are observed under applying electric field although the current peak is registered. However, for this case the switching mechanisms can not be determined. Both, rotation around the cone and rotation around long axis, leaves the optical dielectric tensor eigen directions unchanged.

**Discussion**

Summarizing experimental results, the switching in the $B_{1Rev\ Tilted}$ phase involves either rotation around the long axis or around the tilt cone, and preserves the synclinic structure or leads to the anticlinic ferroelectric state, respectively. In most cases the synclinic state is preferred under electric field, however sterically less favorable anticlinic arrangement of columns could also be induced if the tilt angle is small and the cross section of the column is relatively big, thus the steric hindrances at the column borders have only a minor effect compared to the column bulk. Moreover, it has been observed that even if the switching from the ground synclinic antiferroelectric state to the synclinic ferroelectric state occurs by rotation around the long molecular axis, direct switching between two ferroelectric synclinic states takes place by collective rotation of all molecules on the tilt cone. In this case, upon going around the cone the molecules experience only small energy barrier, as the column boundaries only weakly affect the switching.

To study switching in the $B_{1Rev\ Tilted}$ phase a simple model was considered that takes into account elastic deformations in the director (**n**) field, space deformation of the unit vector in the direction of polarization (**p**) and the electrostatic interactions due to the polarization self-interaction and due to the interaction of polarization with the external electric field (**E**). This is a modified Landau – de Gennes model that has been recently successfully applied to study the director and layer structure in surface stabilized ferroelectric liquid crystals with high spontaneous polarization [16]. Since for the $B_{1Rev\ Tilted}$ phase the tilt angle ($\vartheta$) and the layer normal vector ($\mathbf{v}$) are constant the free energy density can be written as:

$$f = \frac{1}{2}K_n\left[(\nabla\cdot\mathbf{n})^2 + (\nabla\times\mathbf{n})^2\right] + \frac{1}{2}K_p(\nabla\cdot\mathbf{p} - c_0)^2 + \frac{1}{2}K_{np}\left[\mathbf{p}\times(\mathbf{v}\times\mathbf{n})\right]^2 + \frac{P_0^2}{2\varepsilon\varepsilon_0}p_x^2 - P_0\mathbf{p}\cdot\mathbf{E} \qquad (1)$$

The first term in Eq. (1) describes the one-constant ($K_n$) approximation for the elastic deformations (splay and bend) in the director field; the second term describes the splay deformation allowed because polarization in the B-phases is a true vector [17], $c_0$ being a constant stabilizing finite splay of **p**. The elastic constant for polarization deformation is $K_p$. The spontaneous splay of **p** shall not affect the structure in the $B_{1Rev\ Tilted}$ phase significantly, as the width of single column is usually much smaller than the undulation wavelength found in $B_7$ phase [17]. The third term in Eq. (1) with elastic constant $K_{np}$ describes the energy penalty for the polarization vector not being perpendicular to the tilt plane, which is the preferred direction of **p** in the tilted *banana* phases. The last two terms in Eq. (1) describe the dipole self-interaction and the polarization interaction with the external field, respectively. The $P_0$ parameter is the magnitude of the electric polarization and $p_x$ is the $x$-component of the polarization unit vector **p**. In the model it is assumed that the directions of **n** and **p** are coupled, since **p** is preferably perpendicular to **n**. The geometry of the problem is shown in Fig. 1. The column axes are parallel to the $y$-direction and the layer normal is along the $z$-direction. The structure is polarization modulated along the $x$-direction, columns with **p** along $+y$ and $-y$ interchange. The width of a single column is $L$. Upon the switching, the director **n** can rotate on the tilt cone and/or the polarization vector **p** can rotate around the director **n**. The directions of **n** and **p** are determined by the cone (tilt) angle $\vartheta$, azimuthal angle $\varphi$ describing position of director **n** on the tilt cone and the angle $\alpha$, which describes the position of the polarization vector **p** with respect to the tilt plane. The variables $\varphi$ and $\alpha$ change along the $x$-direction only. The proposed model studies spatial variation of **n** and **p** in a single column, since the neighboring columns in $z$ direction have synclinic orientation and switch simultaneously while influence of neighboring columns in $x$ direction is taken into account through the boundary conditions. In the simplest case, when the electric field along the $y$-axis is applied to antiferroelectric structure, the polarization in every second column is parallel to the external field and only columns with polarization anti-parallel to the field will switch. Also, when the field is switched off and the antiferroelectric state is restored, the reorientation of the molecules will occur only in every second column. If switching is such that the molecules in the neighboring column do not rotate, the molecules at the column border respond less readily to the external field. This can be modeled by strong anchoring of **n** and **p** at the column surfaces. However, if the direct switching between two ferroelectric synclinic states takes place (the field sign is suddenly reversed) the molecules in all the columns have to rotate simultaneously. Thus the draw-back on the bordering molecules is significantly smaller, which can be modeled by weak anchoring of **n** and **p** at the column borders.

The type of switching, by rotation around the cone or around the director **n,** is determined by the ratio of the elastic constants, the thickness of the column *L* and the strength of the surface anchoring at the column borders.

Using the definition of the dimensionless free energy *G*, $G = \frac{2\varepsilon\varepsilon_0}{P_0^2 L} F = \int_0^1 g \, d\xi$,

where $F = \int_0^L f \, dx$ and $\xi = x/L$, the dimensionless free energy density *g* can be given as:

$$g = \frac{1}{2}k_n\left[(\nabla_\xi \cdot \mathbf{n})^2 + (\nabla_\xi \times \mathbf{n})^2\right] + \frac{1}{2}k_p(\nabla_\xi \cdot \mathbf{p} - c_0 L)^2 + \frac{1}{2}k_{np}\left[\mathbf{p} \times (\mathbf{v} \times \mathbf{n})\right]^2 + p_x^2 - k_E \mathbf{p} \cdot \hat{\mathbf{e}} \quad , \qquad (2)$$

where the dimensionless parameters are

$$k_n = \frac{2\varepsilon\varepsilon_0 K_n}{P_0^2 L^2}, \quad k_p = \frac{2\varepsilon\varepsilon_0 K_p}{P_0^2 L^2}, \quad k_{np} = \frac{2\varepsilon\varepsilon_0 K_{np}}{P_0^2}, \quad k_E = \frac{2\varepsilon\varepsilon_0 E}{P_0}$$

and $\hat{\mathbf{e}}$ is the unit vector in the direction of the external field $\mathbf{E} = E\hat{\mathbf{e}}$.

The parameter values depend in a complex way on molecular structure however, only two of them $k_n$ and $k_p$ depend on the column width. In materials with large *L,* the $k_n$ and $k_p$ are small compared to $k_{np}$ and switching by rotation on the cone, in which **p** remains perpendicular to the tilt plane, is preferred. This type of switching is observed in a simple lamellar $B_2$ phase for which L is infinite. If *L* is sufficiently small, then $k_n$ and $k_p$ become large compared to $k_{np}$ and switching by rotation around the long molecular axis is preferred, which, indeed, is commonly observed in $B_{1Rev\ Tilted}$ phase. If all the parameters are of the same order of magnitude the switching is a complex combination of both rotation around the long axis and the rotation on the cone.

To find the director and polarization vector field in one column at a certain time $t$ after the field was switched on/off or reversed, the simplest approach [18], i.e. Landau-Khalatnik equations was used:

$$\frac{\partial g}{\partial \alpha} - \frac{d}{d\xi}\frac{\partial g}{\partial \alpha_\xi} + \frac{2\varepsilon\varepsilon_0\eta}{P_0^2}\frac{\partial \alpha}{\partial t} = 0$$

$$\frac{\partial g}{\partial \varphi} - \frac{d}{d\xi}\frac{\partial g}{\partial \varphi_\xi} + \frac{2\varepsilon\varepsilon_0\eta}{P_0^2}\frac{\partial \varphi}{\partial t} = 0$$

For simplicity it was assumed that the effective viscosities ($\eta$) for the rotation around the long molecular axis and around the cone are the same.

The results of the numerical calculations are given in Figs. 6 and 7. In Fig. 6 the spatial dependence for **n** and **p** across the column in equilibrium state, which was obtained by applying external electric field to the initial antiferroelectric synclinic state, is shown. At $k_n$ = 10 ($L$=5 nm, which is rather typical value in $B_{1Rev\ Tilted}$ phase) the switched-on state occurs by rotation around the long axis and at $k_n$ = 0.1 ($L$=50 nm) the column switches by rotation on the cone. As discussed before, strong anchoring at the column borders was assumed ($\alpha(0) = \alpha(L) = 0$ and $\varphi(0) = \varphi(L) = 0$). To find the type of switching when the external field is reversed, surface anchoring at the column border is released, since in this case molecules in all columns have to rotate simultaneously. Thus the column boundary is approximated by requiring $d\alpha/\mathrm{d}x = 0$ and $d\varphi/dx = 0$ both at $x$=0 and $x$=$L$. In this case the rotation around the cone is always observed independently of the column size. In Fig. 7 the time dependence of $\varphi_c$ and $\alpha_c$ for the molecules in the center of the column ($x$=$L$/2) is shown for the case of small column cross-section. At the initial stage of switching both $\varphi$ and $\alpha$ angles change simultaneously, however the final result depends on the external field. When the field is reversed the structure switches between two ferro-, synclinic states that differ by $\pi$ in the azimuthal cone angle $\varphi$. However, when the field is turned-off, the structure relaxes from ferroelectric, synclinic state back to the antiferro, synclinic state that differ by $\pi$ in the polarization direction, denoted by $\alpha$.

Finally, the dielectric response as a function of the column width has been estimated. The eigenvalues of the susceptibility matrix, i.e. the matrix of the second derivatives of the dimensionless free energy with respect to the variables $\alpha$ and $\varphi$, were calculated according to the procedure described by Čopič *et al.* [19]. The matrix elements were evaluated in the equilibrium state without the external field ($\varphi(x)=0$ and $\alpha(x)=0$). The lowest eigenvalue ($\lambda$) of the susceptibility matrix is proportional to the relaxation frequency of the main mode with some effective viscosity as the proportionality factor. Numerical calculations show that this eigenvalue $\lambda$ increases, and consequently the corresponding mode relaxation moves to higher frequencies, with decreasing column thickness $L$. At large $L$ the main contribution to the eigenvector which corresponds to the lowest frequency mode comes from the phason fluctuations $\delta\varphi$, while polarization fluctuations $\delta\alpha$ are rather small. If $L$ is reduced than both fluctuations become comparable. At very small $L$ the fluctuations $\delta\alpha$ become dominant. Usually the magnitude of the dielectric response, $\Delta\varepsilon$, is determined mainly by the lowest eigenvalue of the susceptibility matrix, and it is inversely proportional to $\lambda$. The numerical results suggest that the dielectric response decreases with decreasing $L$. This agrees with the experiment which shows, that the dielectric response in the $B_2$ phase (limit of infinite $L$) is considerably stronger and has lower relaxation frequency than in the $B_1$ – type phase.

**Acknowledgements**

This work was supported by the KBN grant 4T09A 00425.

Table 1. Chemical formulas, phase transition temperatures (°C) and enthalpy changes (in the parentheses, J $g^{-1}$) for the studied materials. The parameters of the crystallographic unit cells are given in Å. $B_{1R}$ and $B_{1RT}$ were used for short, they stand for $B_{1Rev}$ and $B_{1Rev\ Tilted}$, respectively.

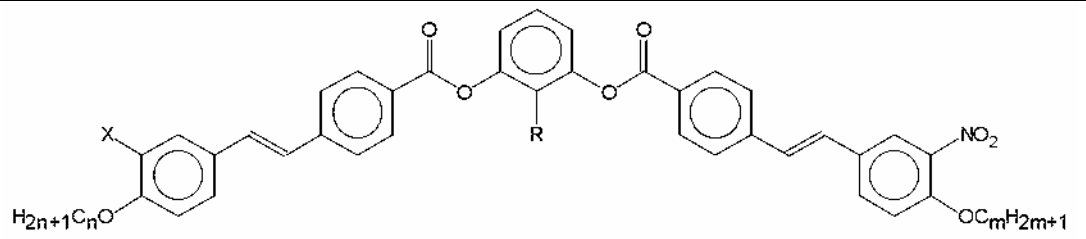

| | R | X | n | m | phase sequence<br>crystallographic unit cell parameters |
|---|---|---|---|---|---|
| **1** | $CH_3$ | Cl | 14 | 14 | Cry 116.5 (22.1) $B_{1RT}$ 127.7 (1.0) $SmAP_R$ 141.3 (9.1) Iso<br>$B_{1RT}$: a = 214.4, b = 51.6, γ = 82° |
| **2** | $CH_3$ | I | 12 | 12 | Cry 89.2 (31.7) $B_{1RT}$ 112.9 (1.0) $SmAP_R$ 125.6 (8.6) Iso<br>$B_{1RT}$ : a = 112, b = 51.3, γ = 84° |
| **3** | $CH_3$ | $OCH_3$ | 12 | 12 | Cry 98.8 (20.1) $B_{1RT}$ 106.2 (0.9) $SmAP_R$ 117.6 (10.3) Iso<br>$B_{1RT}$ : a = 210.0, b = 48.4, γ = 81.5° |
| **4** | H | Cl | 12 | 12 | Cry 113.6 (13.2) $B_{1RT}$ 146.9 (17.5) Iso<br>$B_{1RT}$ : a = 178.0, b = 45.7, γ = 77° |
| **5** | H | I | 16 | 12 | Cry 111.1 (3.7) $B_{1RT}$ 129.7 (2.3) $SmAP_R$ 136.6 (7.5) Iso<br>$B_{1RT}$ : a = 192, b = 46.9, γ = 75° |

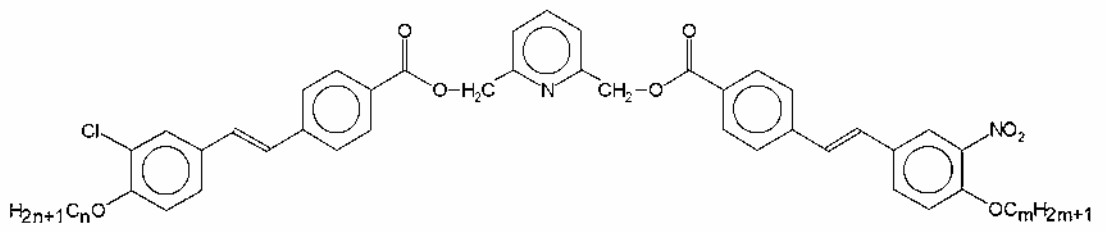

| | n | m | |
|---|---|---|---|
| **6** | 12 | 12 | Cry 118.0 (2.0) $B_{1RT}$ 138.4 (2.3) $B_{1R}$ 143.7 (21.8) Iso<br>$B_{1R}$ : a = 140.0, b = 53.0<br>$B_{1RT}$ : a = 104.2, b = 51.8, γ = 79° |
| **7** | 12 | 14 | Cry 116.4 (1.8) $B_{1RT}$ 138.6 (2.8) $B_{1R}$ 145.6 (21.6) Iso<br>$B_{1R}$ : a = 139.0, b = 53.2<br>$B_{1RT}$ : a = 102.7, b = 52.1, γ = 78° |

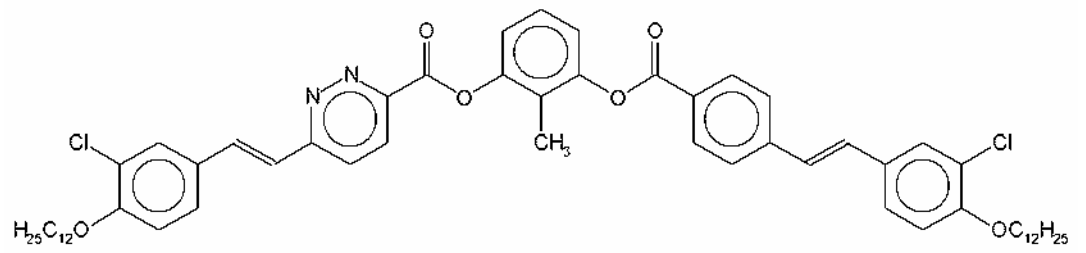

| | |
|---|---|
| **8** | Cry 130.8 (65.2) $B_2$ 132.0 (11.4) Iso |

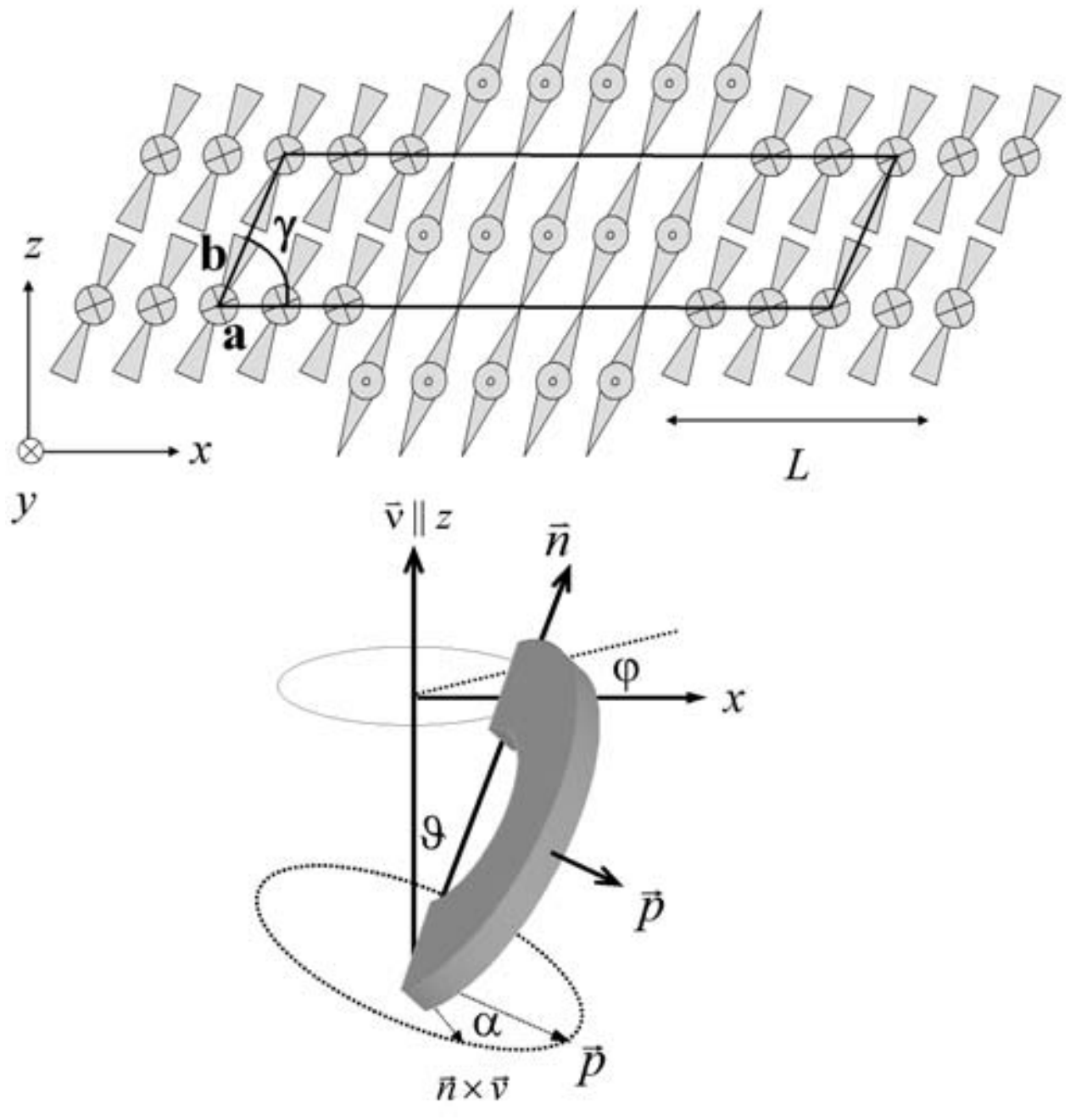

Fig. 1. Schematic drawing of the $B_{1Rev\ Tilted}$ phase structure, *L* is the width of the column cross-section. The column axes are parallel to the *y*-direction and the layer normal (ν) is along the *z*-direction. The structure is the polarization modulated along the *x*-direction. Orientation of single *banana* molecule in the $B_{1Rev\ Tilted}$ phase: the polarization unit vector **p** is perpendicular to the long molecular axis and director **n** is along long molecular axis. The orientation of **p** and **n** vectors is determined by the cone angle ϑ, azimuthal angle φ describing position of the director **n** on the cone, and the angle α, which describes the position of the polarization unit vector **p** with respect to the tilt plane normal defined by **n**×ν.

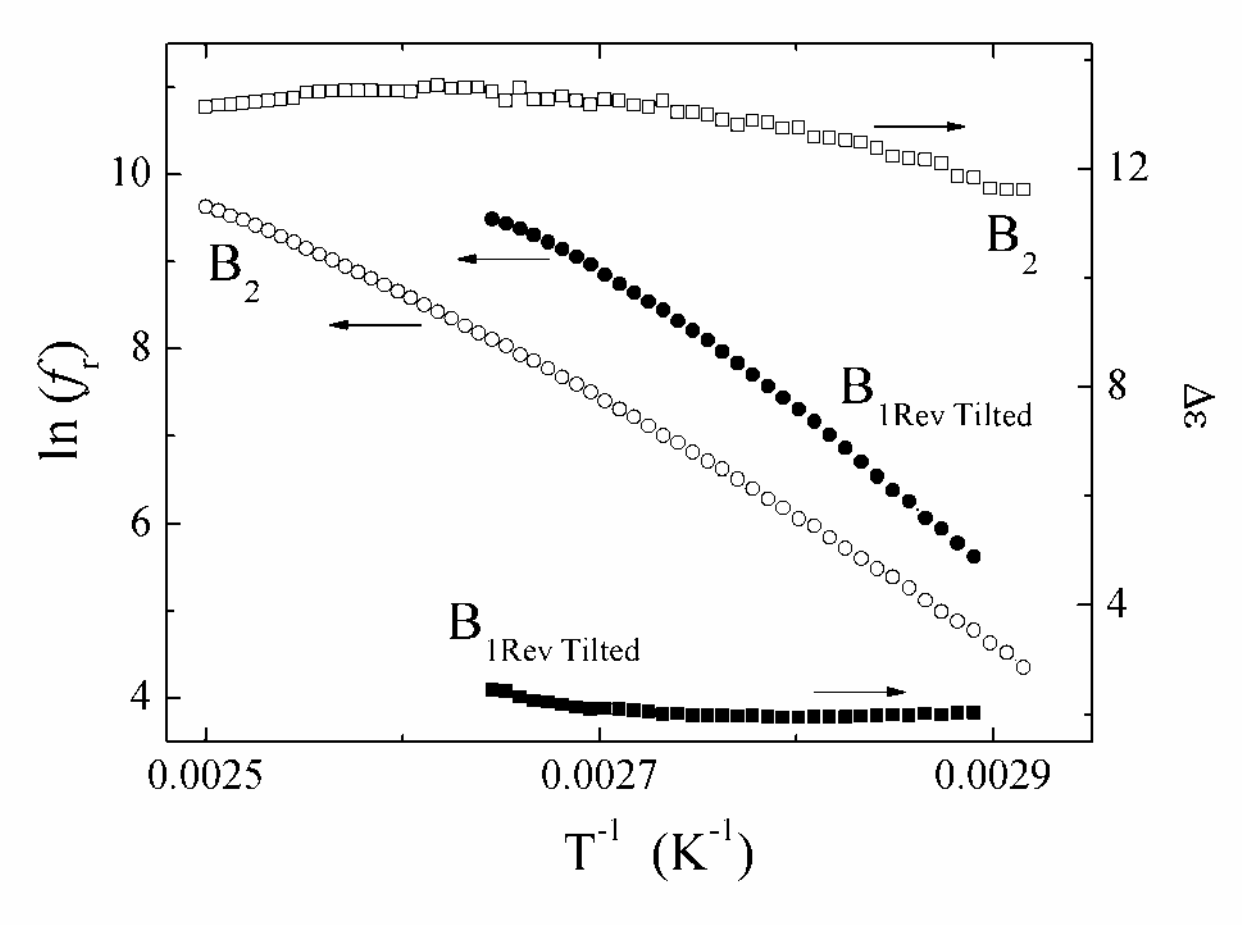

Fig 2. Dielectric response – the inverse temperature dependence of the relaxation frequency $f_r$ (circles) and the mode strength Δε (squares) in the $B_{1Rev\ Tilted}$ phase of compound **3** (solid symbols) and in the lamellar $B_2$ phase of the compound **8** (open symbols). Dielectric permittivity dispersion was measured on cooling in a 5 μm thick cells and analyzed using the Cole-Cole equation. The activation energy obtained from the fitting of $f_r$ to the Arhenius law gives $E_a \sim 145$ kJ mol$^{-1}$ and 110 kJ mol$^{-1}$ for the $B_{1Rev\ Tilted}$ and the $B_2$ phase, respectively.

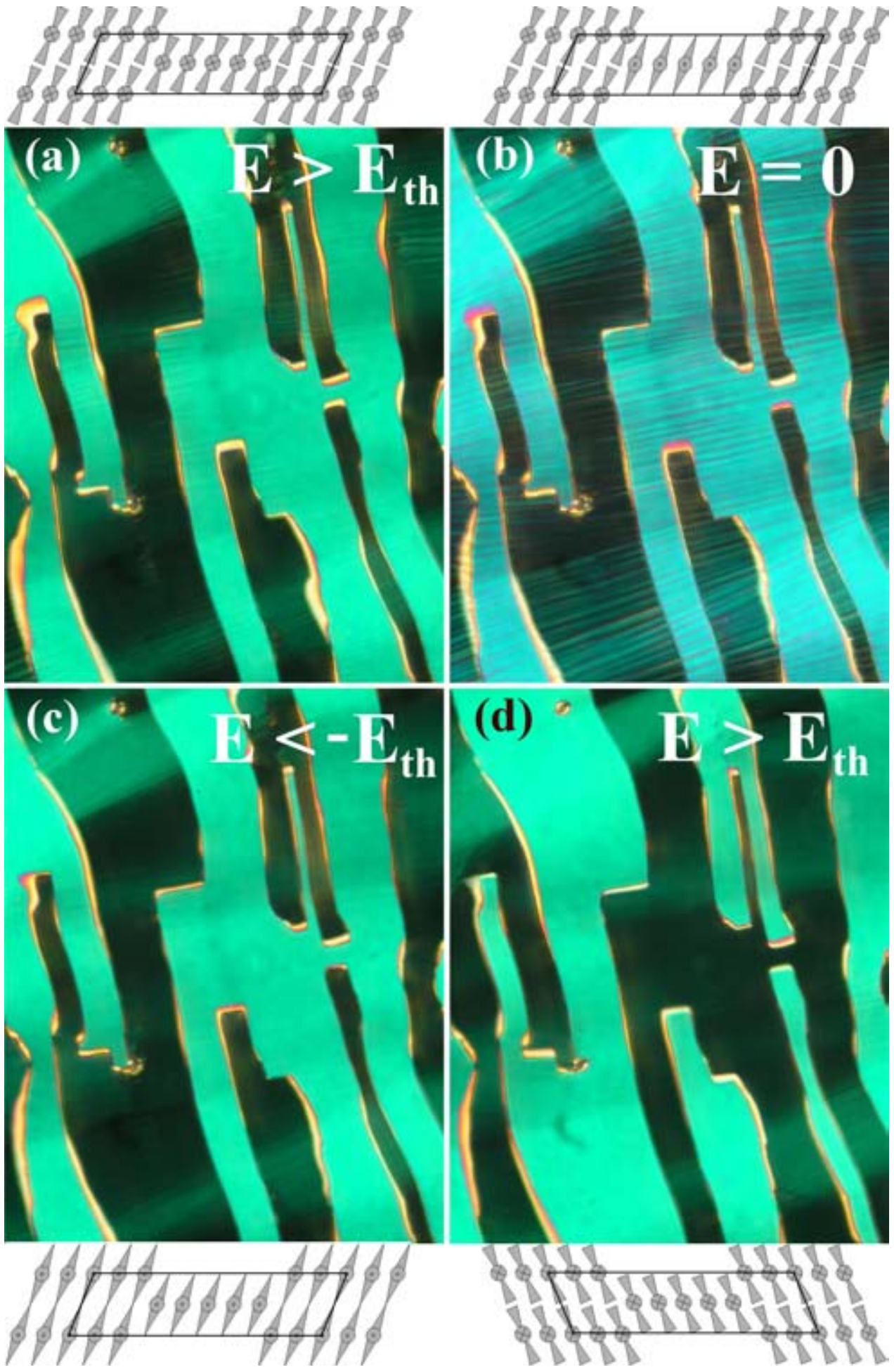

Fig. 3. Texture of the $B_{1Rev\ Tilted}$ phase of compound **1**, electric field is applied in sequence +E, 0, -E, +E (figs. a-d, respectively). Reversion of the tilt direction is observed when the field sign is suddenly reversed (c)-(d). Schematic drawing of molecular orientation in crystallographic unit cell is also given; however note that the real structure involves also splay deformation of polarization vectors inside the columns (see the text for explanation).

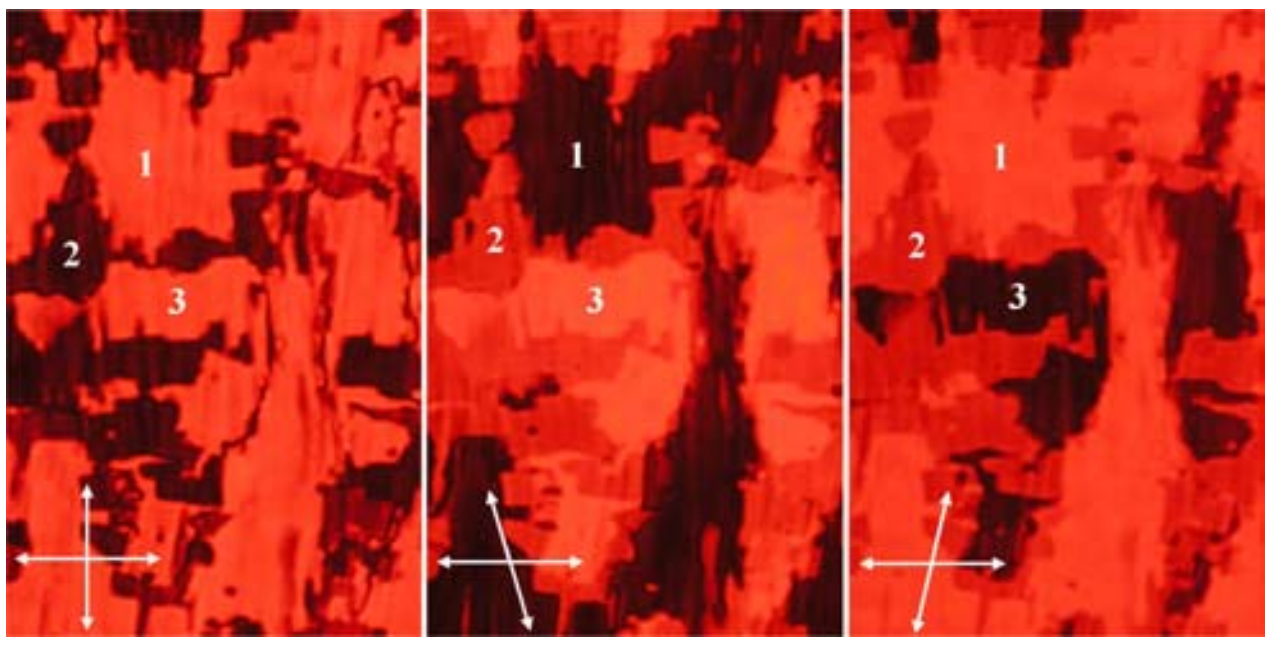

Fig. 4. Texture of the $B_{1Rev\ Tilted}$ phase of compound **1** with optically active regions at the boundaries of synclinc domains, viewed with a red filter under crossed and slightly de-crossed polarizes. Regions marked as 1 and 3 are optically active with the opposite handedness, while region 2 does not show optical activity. The red filter was used to avoid color effects due to birefringence. Optically active regions could be significantly suppressed in course of applying electric field, see yellow domain boundaries in Fig. 3.

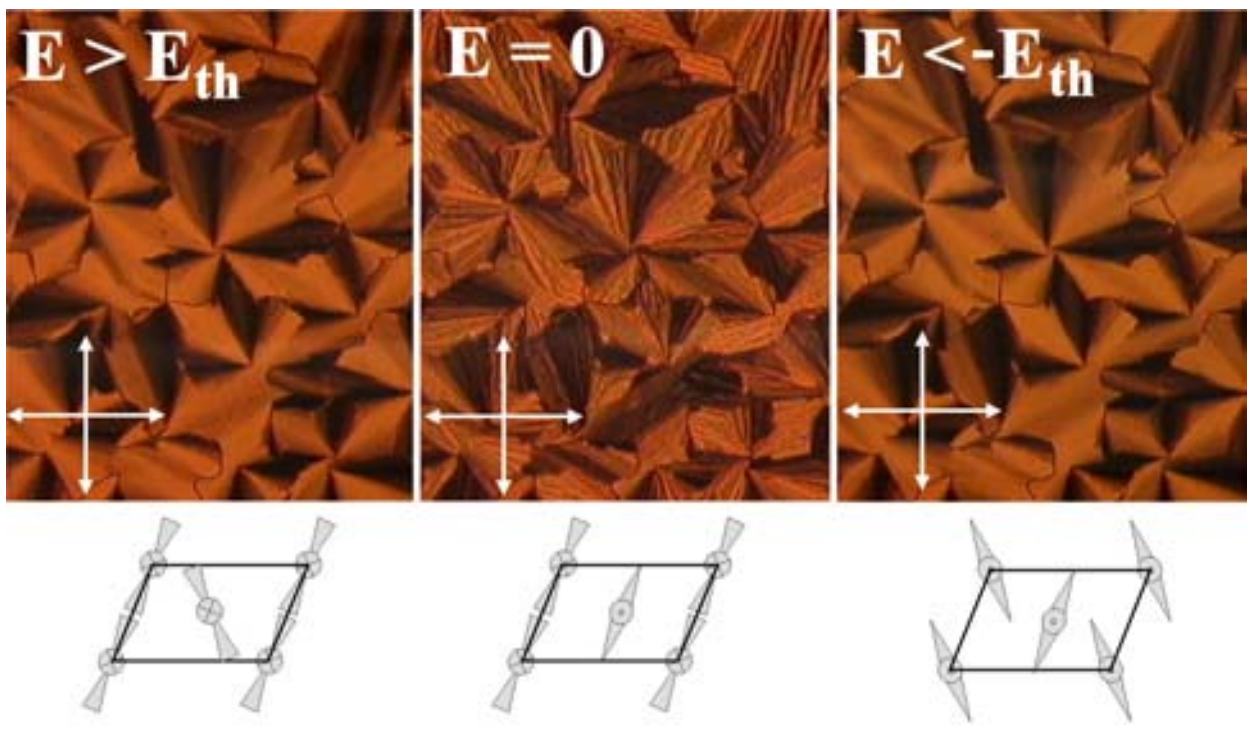

Fig . 5. Texture in the $B_{1Rev\ Tilted}$ phase of the compound **7**. In the field-off state synclinic domains are observed, application of the electric field induces anticlinic structures. The schematic drawing of molecular orientation in the crystallographic unit cell is given bellow photos. For simplicity of the figure only the orientation of molecules in the middle of each column is drawn, space-filling model requires bend-splay deformation of director inside column (see the text for explanation).

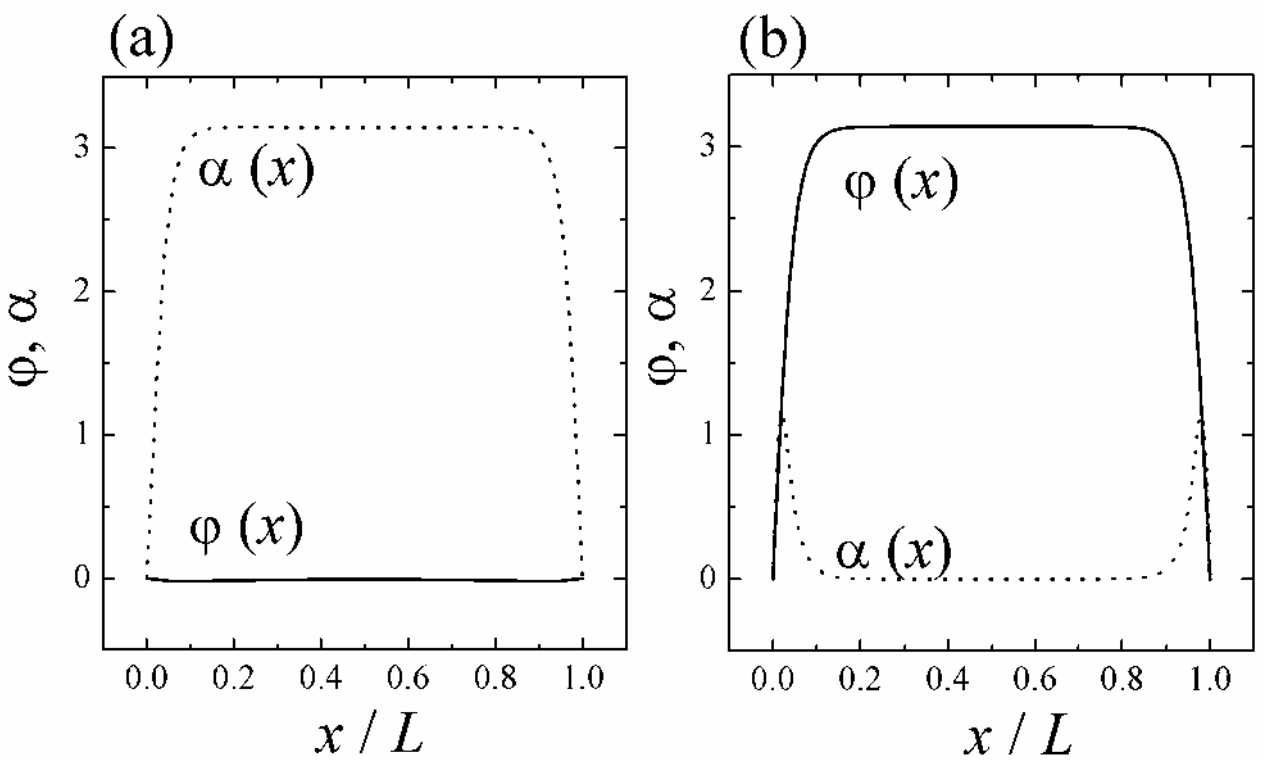

Fig. 6. Calculated director position on the cone, φ (solid line) and the direction of the polarization, α (dotted line) in the switched–on state, in the case of (a) small $L$, $k_n$=10 and (b) large $L$, $k_n$=0.1. The initial structure in the column, without external field was $\alpha(x)$=0 and $\varphi(x)$=0. The set of parameters used for the calculation was: $k_{np}$ = 200 and $k_p$ = 0.01 $k_n$, $\vartheta$=10°, $P_0$ = 100 nC/cm$^2$, ε = 10, $K_n$ = $10^{-13}$ N, η ≈ 0,1 kg/(ms), $\tau = 2\varepsilon\varepsilon_0\eta / P_0^2$, $c_0 = 2\pi/d_s$ where $d_s$ ~30 nm [17].

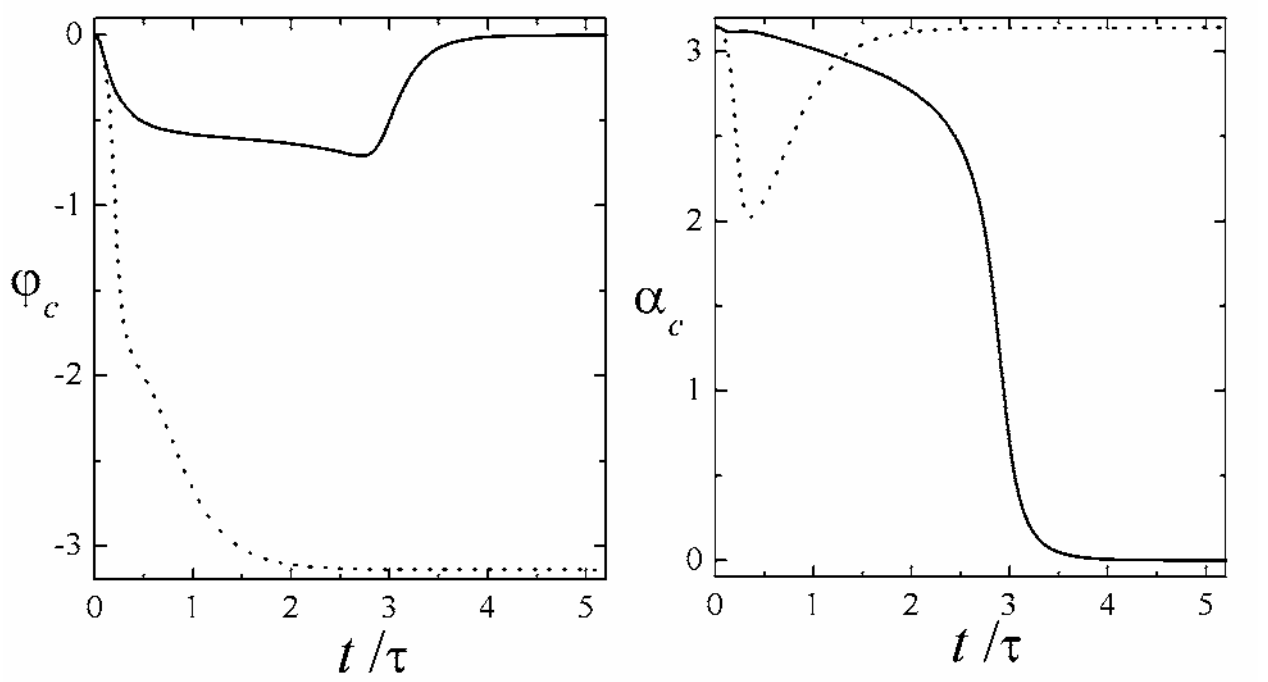

Fig. 7. Time dependence of $\varphi_c$ and $\alpha_c$ in the center of the column (small $L$ regime). The initial structure was the switched-on state shown in Fig. 6a. At $t$=0 the field was switched off (solid lines) or reversed (dotted lines). The set of parameters used for the calculation was: $k_n$=10, $k_{np}$ = 200 and $k_p$ = 0.01 $k_n$, $k_E$ =-10 (in the case of the reversed field), $\vartheta$=10°, $P_0$ = 100 nC/cm$^2$, ε = 10, $K_n$ = $10^{-13}$ N and η ≈ 0,1 kg/(ms), $\tau = 2\varepsilon\varepsilon_0\eta / P_0^2$, $c_0 = 2\pi/d_s$ and $d_s$ ~30 nm [17].